\documentclass{pj}
\usepackage[pdftex]{graphicx}
\usepackage{amssymb}
\usepackage{hyperref}[2003/11/30]
\usepackage{color}
\usepackage{verbatim}
\usepackage{amsmath}
\usepackage{upgreek}
\usepackage{epstopdf}

\begin{document}
\setcounter{page}{1}
\pjheader{Vol.\ x, y--z, 2004}
%\jheader{Vol.\ x, y--z, 2004}
%\pheader{Vol.\ x, y--z, 2004}

\title[Permittivity measurements of biological samples\dots]{Permittivity measurements of biological samples by an open-ended coaxial line}

%Authors:   Jake Bobowski	      Jake.Bobowski@ubc.ca
%		    Thomas Johnson	      Thomas.Johnson@ubc.ca
%
%Corresponding Author:  Jake Bobowski   Jake.Bobowski@ubc.ca

\author{J.~S.~Bobowski and T.~Johnson}

\address{School of Engineering, University of
British Columbia Okanagan\\
3333 University Way, Kelowna, British
Columbia V1V 1V7, Canada}

\runningauthor{Bobowski}
\tocauthor{J.~Bobowski}

\begin{abstract}
We previously reported on the complex permittivity and dc conductivity of waste-activated sludge.  The measurements, spanning a frequency range of 3~MHz to 40~GHz, were made using an open-ended coaxial transmission line.  Although this technique is well established in the literature, we found that it was necessary to combine methods from several papers to use the open-ended coaxial probe to reliably characterize biological samples having a high dc conductivity.  Here, we provide a set of detailed and practical guidelines that can be used to determine the permittivity and conductivity of biological samples over a broad frequency range.  Due to the electrode polarization effect, low frequency measurements of conducting samples require corrections to extract the intrinsic electrical properties.  We describe one practical correction scheme and verify its reliability using a control sample.

\end{abstract}

\section{Introduction}\label{sec:intro}

As a versatile tool for measuring the real and imaginary components of the permittivity of materials, the open-ended coaxial probe has found widespread use among researchers spanning numerous disciplines.  Examples of materials characterized using the open-ended coaxial probe include, but are not limited to: biological tissues~\cite{Stuchly:1982}, tumors~\cite{Foster:1981}, binary mixtures of liquids~\cite{Bao:1996}, particle suspensions and emulsions~\cite{Erle:2000}, food~\cite{Wang:2003}, vegetation~\cite{El-Rayes:1987}, and soil~\cite{Jackson:1990}.  The advantages of the open-ended coaxial probe over other techniques are that it is a broadband measurement ($10^5$--$10^{10}$~Hz), requires no sample preparation, and is suitable for liquid and semisolid samples~\cite{Kaatze:2006}.

Using the open-ended coaxial probe, we have made the first measurements of the complex permittivity and conductivity of thickened waste-activated sludge (WAS)~\cite{Li:1990} sampled from our local wastewater treatment facility (WWTF)~\cite{Bobowski:2012}.  Characterizing the electrical properties of materials has both scientific and practical value.  For example, our measurements of WAS were motivated by the practical desire to optimize the electromagnetic pretreatments of WAS.  These pretreatments can be used prior to anaerobic digestion to enhance the production rate of biogas during the digestion stage~\cite{Eskicioglu:2009,Appels:2008}.  Although the electrical properties of various biological materials have been studied for many decades, a complete understanding of the microscopic mechanisms leading to the observed dielectric dispersions has not yet been achieved and are typically modeled using empirical results~\cite{Pethig:1984}.  Studies that further our understanding of these mechanisms are, therefore, of fundamental interest.

The goal of this work is twofold:  First, we present a concise summary of the calibration methods used to make accurate permittivity measurements using open-ended coaxial probes.  Second, we demonstrate the capabilities and limitations of the measurement technique using two control samples and a determination of the unknown complex permittivity and conductivity of WAS.

As described in Sec.~\ref{sec:prelim}, the experimental method consists of sending an incident signal down a length of semi-rigid coaxial cable whose open end is submerged in the material under test (MUT).  In Sec.~\ref{sec:imped}, we discuss the effective impedance of the submerged probe tip and show that it is sensitive to the surrounding material.  The signal reflected at the probe tip is measured and analyzed to determine the electrical properties of the MUT.  At frequencies below 100~MHz, the coaxial probe is treated as an ideal transmission line and the permittivity and conductivity of the MUT are directly related to the measured reflection coefficient.  We describe this case in Sec.~\ref{sec:probe} and investigate the limits of this analysis using methyl alcohol and saltwater control samples.  At higher frequencies, both ohmic losses in the probe conductors and dielectric losses in the probe insulator become non-negligible.  Additionally, spurious reflections at the probe connector can occur.  Section~\ref{sec:analysis} introduces a calibration scheme that corrects the measured reflection coefficient for these effects.  The scheme uses a set of three standard terminations (open, short, and known load) at the open end of the probe.  The methyl alcohol and saltwater control samples are again used to evaluate the performance of the calibration scheme from low frequency up to 40~GHz.  For highly conducting samples, the submerged end of the coaxial probe acquires a surface charge at sufficiently low frequencies.  This surface charge alters the effective impedance of the probe tip and hence the measured the reflection coefficient.  Section~\ref{sec:electrodePolarization} presents a reliable technique to identify and then correct for the systematic errors introduced by the electrode polarization effect.  The correction scheme is applied to the saltwater control sample to demonstrate its reliability.  Newly-determined WAS permittivity data, along with a brief discussion of the relevant dispersion mechanisms, are presented in Sec.~\ref{sec:WAS}. A summary of the key conclusions is given in Sec.~\ref{sec:conclusions}.

\section{Experimental Geometry}\label{sec:prelim}

When a material is exposed to a time-harmonic electric field $\mathbf{E}e^{j\omega t}$ of angular frequency $\omega$, the total current density $\mathbf{J}$ is the sum of the conduction and displacement current densities:
\begin{equation}
\mathbf{J}=\left(\sigma_{dc}+j\omega\varepsilon_0\varepsilon_r\right)\mathbf{E}=\left[\left(\sigma_{dc}+\omega\varepsilon_0\varepsilon^{\prime\!\prime}\right)+j\omega\varepsilon_0\varepsilon^\prime\right]\mathbf{E},
\end{equation}
where $\varepsilon_0$ is the permittivity of free space, $\sigma_{dc}$ is the dc conductivity, and \mbox{$\varepsilon_r=\varepsilon^\prime-j\varepsilon^{\prime\!\prime}$} is the relative permittivity of the material.  As it is not possible to separate the contributions of $\sigma_{dc}$ and $\varepsilon^{\prime\!\prime}$ in an experimental measurement, many authors choose to define a frequency-dependent electrical conductivity $\kappa(\omega)\equiv\sigma_{dc}+\omega\varepsilon_0\varepsilon^{\prime\!\prime}$.

\begin{figure}[tb]
\begin{center}
\includegraphics[width=0.85\columnwidth]{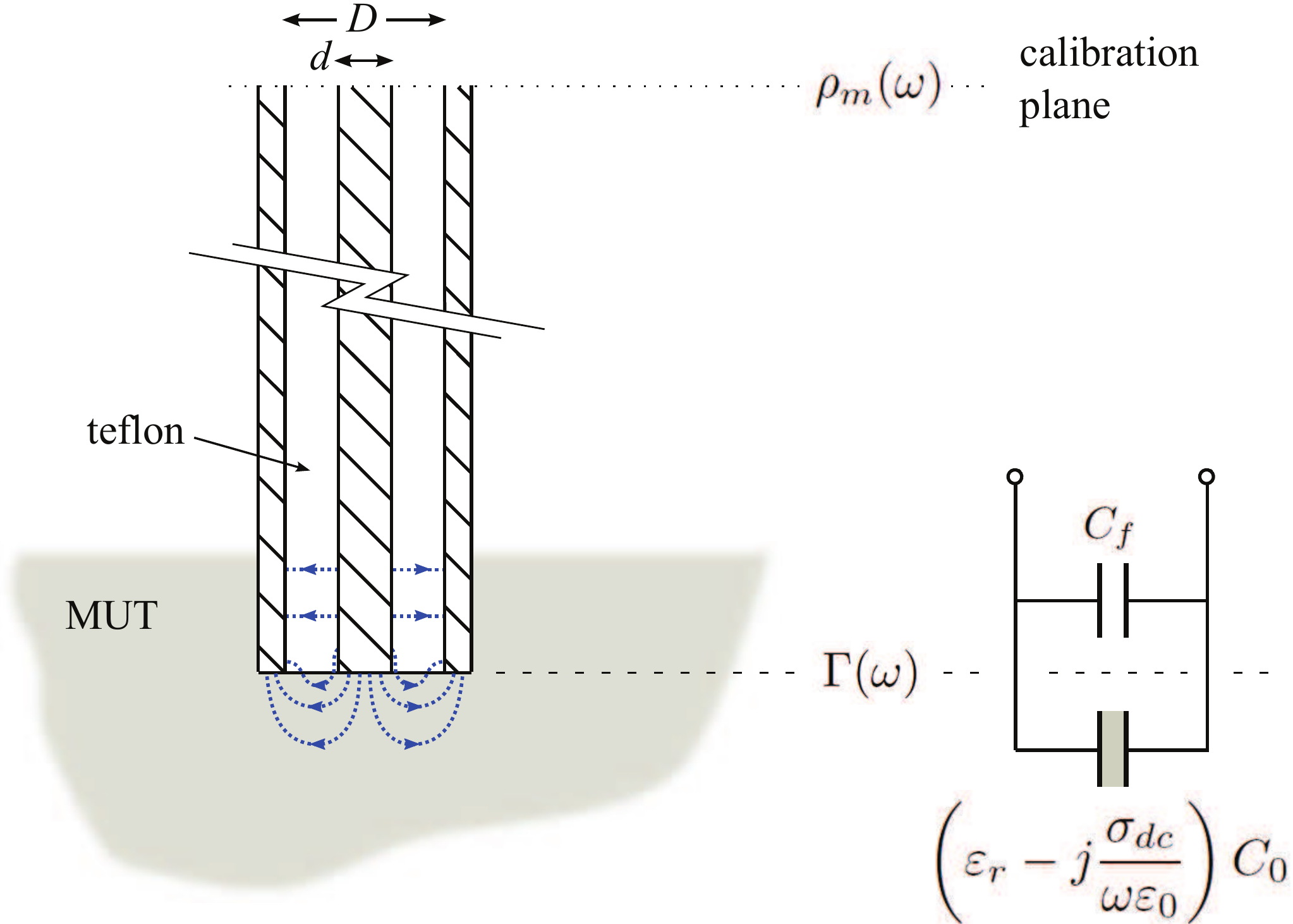}
\caption{\label{fig:fringe}A schematic of the experimental geometry showing the open end of the coaxial probe submerged in the MUT.  The calibration plane of the VNA is established at the opposite end of the probe.  The reflection coefficient $\rho_m(\omega)$ at the calibration plane is measured directly by the VNA.  In sections \ref{sec:probe} and \ref{sec:analysis}, $\rho_m(\omega)$ is related to the reflection coefficient $\Gamma(\omega)$ at the probe tip which can be used to determine the permittivity $\varepsilon_r$ and conductivity $\sigma_{dc}$ of the MUT.  The dotted lines represent electric field lines in the vicinity of the interface between the probe tip and the MUT.  The impedance of the interface is modeled as parallel shunt capacitances.  $C_f$ accounts for fringing of the electric field that occurs within the teflon dielectric of the coaxial probe and $\left(\varepsilon_r-j\sigma_{dc}/\omega\varepsilon_0\right)C_0$ for the fringing fields in the MUT.}
\end{center}
\end{figure}

The goal of this work is to describe practical data acquisition and data analysis methods that can be used to reliably characterize the electrical properties of materials having appreciable dc conductivity.  All measurements are made using open-ended coaxial probes fabricated from short sections of commercially-available semirigid coaxial transmission lines.  One end of the probe was fitted with a connector and the opposite end was polished flat creating an open circuit condition.  During the measurements, the open end of the probe is submerged into the MUT resulting in a mismatched load.  Incident signals, generated using a vector network analyzer (VNA), are partially reflected at the MUT-probe interface and, by analyzing the amplitude and phase of the reflection coefficient as a function of frequency, the complex permittivity and dc conductivity of the MUT can be determined.  The experimental geometry is shown in Fig.~\ref{fig:fringe}.

The next two sections explicitly show how $\varepsilon_r$ and $\sigma_{dc}$ are extracted from the measured reflection coefficient $\rho_m(\omega)$.  We start by treating the low-frequency case, in which the probe is modeled as an ideal transmission line.  From this analysis, we obtain a simple relationship between the measured reflection coefficient $\rho_m$ at the VNA calibration plane and the desired reflection coefficient $\Gamma$ at the probe tip.  This section also shows how to characterize the effective impedance of the open end of the probe when it is submerged in a material of permittivity $\varepsilon_r$ and conductivity $\sigma_{dc}$.  Section~\ref{sec:analysis} considers the high-frequency case where losses along the length of the probe cannot be neglected.  The probe is modeled as a two-port network such that $\rho_m$ and $\Gamma$ can be related through a scattering matrix.  A calibration procedure to determine the elements of the scattering matrix is described.

\section{Effective Impedance of Probe Tip}\label{sec:imped}

Below the cutoff frequency of the coaxial transmission line, the transverse electromagnetic (TEM) mode is the only propagating mode.  In this mode, the electric field is radial and the magnetic field encircles the center conductor everywhere inside the coaxial probe except near the open tip where fringing occurs due to the abrupt change in impedance.  This effect is typically modeled by representing the load impedance at the probe tip as two parallel shunt capacitors $C_f$ and $C(\varepsilon_r,\sigma_{dc})$. Here, $C_f$ is purely reactive and accounts for fringing that occurs within the dielectric of the coaxial probe (assumed lossless) whereas $C(\varepsilon_r,\sigma_{dc})=\left(\varepsilon_r-j\sigma_{dc}/\omega\varepsilon_0\right)C_0$ accounts for fringing that occurs within the MUT (see Fig.~\ref{fig:fringe})~\cite{Stuchly:1980}.  Within this model, the load admittance at the probe tip is:
\begin{equation}
Y_L=C_0\left(\omega\varepsilon^{\prime\!\prime}+\frac{\sigma_{dc}}{\varepsilon_0}\right)+j\omega\left(C_f+\varepsilon^\prime C_0\right).\label{eq:Zinv}
\end{equation}
In practice, this simple description of the electromagnetic field distribution at the open end of the probe breaks down well before cutoff.  This failure is due to the onset of evanescent modes at the probe tip.  In a cylindrical coaxial cable with relative dielectric constant $\varepsilon$, the first mode to propagate (other than the TEM mode) is the TE$_{11}$ mode with a cutoff frequency of $f_c\approx 2c/\sqrt{\varepsilon}\pi(D+d)$ where $d$ is the diameter of the center conductor and $D$ is the inside diameter of the outer conductor.  For biological samples at radio and microwave frequencies, a strong suppression of $f_c$ at the junction between the probe tip and the MUT is anticipated since the condition $|\varepsilon_r|\gg 1$ is typically satisfied.

In Ref.~\cite{Jarvis:1994}, Baker-Jarvis {\it et al.\/} show a more complete multimode analysis of electromagnetic fields at the probe tip in the MUT. Numerical methods are used to improve the accuracy of modeling modes at the coaxial tip. However, the method is more complex, and in this work we use the shunt capacitance model as described above to obtain accurate measurements over a wide range of frequencies. The latter method also has the advantage of simplifying the extraction of MUT permittivity from  a set of measured reflection coefficients ($\rho_m$) using analytic equations. The work of Baker-Jaris {\it et al.\/} also includes an analysis of sub-millimeter gaps between the probe tip and the MUT. These so-called ``lift-off'' effects are quantitatively characterized in Ref.~\cite{Jarvis:1994}. However, most biological samples are well characterized without any gap because they are in the liquid or semisolid state or composed of soft tissue. In this work, the probe tip was fully immersed in the sample and no gap was introduced in the measurement method.

\section{Low-Frequency Analysis}\label{sec:probe}

This section considers the low-frequency limit where the ohmic and dielectric losses of the coaxial probe are neglected and the connector used at the opposite end of the probe is assumed to be perfectly transmitting.  First, an expression relating the measured reflection coefficient $\rho_m$ and the reflection coefficient at the probe tip $\Gamma$ is developed.  In subsection~\ref{sec:betaell}, a short circuit at the probe tip is used to determine the propagation constant of the coaxial probe and to evaluate the frequency range over which the probe may be suitably approximated as an ideal transmission line.  Lastly, subsection~\ref{sec:CfC0} uses two liquid samples of known permittivity to experimentally determine $C_f$ and $C_0$ of a particular open-ended coaxial probe.

The total impedance presented by a lossless coaxial line of length $\ell$ is given by:
\begin{equation}
Z_{in}(\omega)=Z_0\frac{1+jZ_0Z_{L}^{-1}\tan\left(\beta\ell\right)}{Z_0Z_{L}^{-1}+j\tan\left(\beta\ell\right)}
\end{equation}
where $Z_0$ is the characteristic impedance of the transmission line, $Z_{L}^{-1}=Y_{L}$, and $\beta$ is the frequency-dependent propagation constant of the probe dielectric.  The probe impedance $Z_{in}$ is related to the measured reflection coefficient $\rho_m$ at the calibration plane of the VNA via:
\begin{equation}
\rho_{m}(\omega)=\frac{Z_{in}-Z_0}{Z_{in}+Z_0}.
\end{equation}
The reflection coefficient at the probe tip $\Gamma(\omega)$ is determined from the effective impedance of the probe tip $Z_L$:
\begin{equation}
\Gamma(\omega)=\frac{Z_{L}-Z_0}{Z_{L}+Z_0},\label{eq:Gamma1}
\end{equation}
and is related to $\rho_{m}(\omega)$ through:
\begin{equation}
\Gamma(\omega)=\rho_{m}(\omega)\frac{1+j\tan\beta\ell}{1-j\tan\beta\ell}.\label{eq:Gamma2}
\end{equation}

\subsection{Determining $\beta\ell$}\label{sec:betaell}

Extracting $Z_{L}$ from measurements of $\rho_{m}(\omega)$ requires knowledge of $\beta\ell$.  For a lossless dielectric, the propagation constant is $\beta=\omega\sqrt{\varepsilon}/c$ where $\varepsilon$ is the dielectric constant of the probe dielectric and $c$ is the free-space speed of light.  One method of determining $\sqrt{\varepsilon}\ell/c$ is by short circuiting the center conductor of the coaxial transmission line to its outer conductor at the probe tip.  High-quality and repeatable short circuits can be achieved by pressing the probe tip into a thin sheet of indium metal.  In this case, $Z_{L}=0$, $\Gamma=-1$, and:
\begin{equation}
\rho_{m}(\omega)=\frac{j\tan\beta\ell-1}{j\tan\beta\ell+1}.\label{eq:rho}
\end{equation}
This expression can be fitted to the measured reflection coefficient of the short-circuited probe with $\sqrt{\varepsilon}\ell/c$ being the only fit parameter.

\begin{figure}[tb]
\begin{center}
\includegraphics[width=0.7\columnwidth]{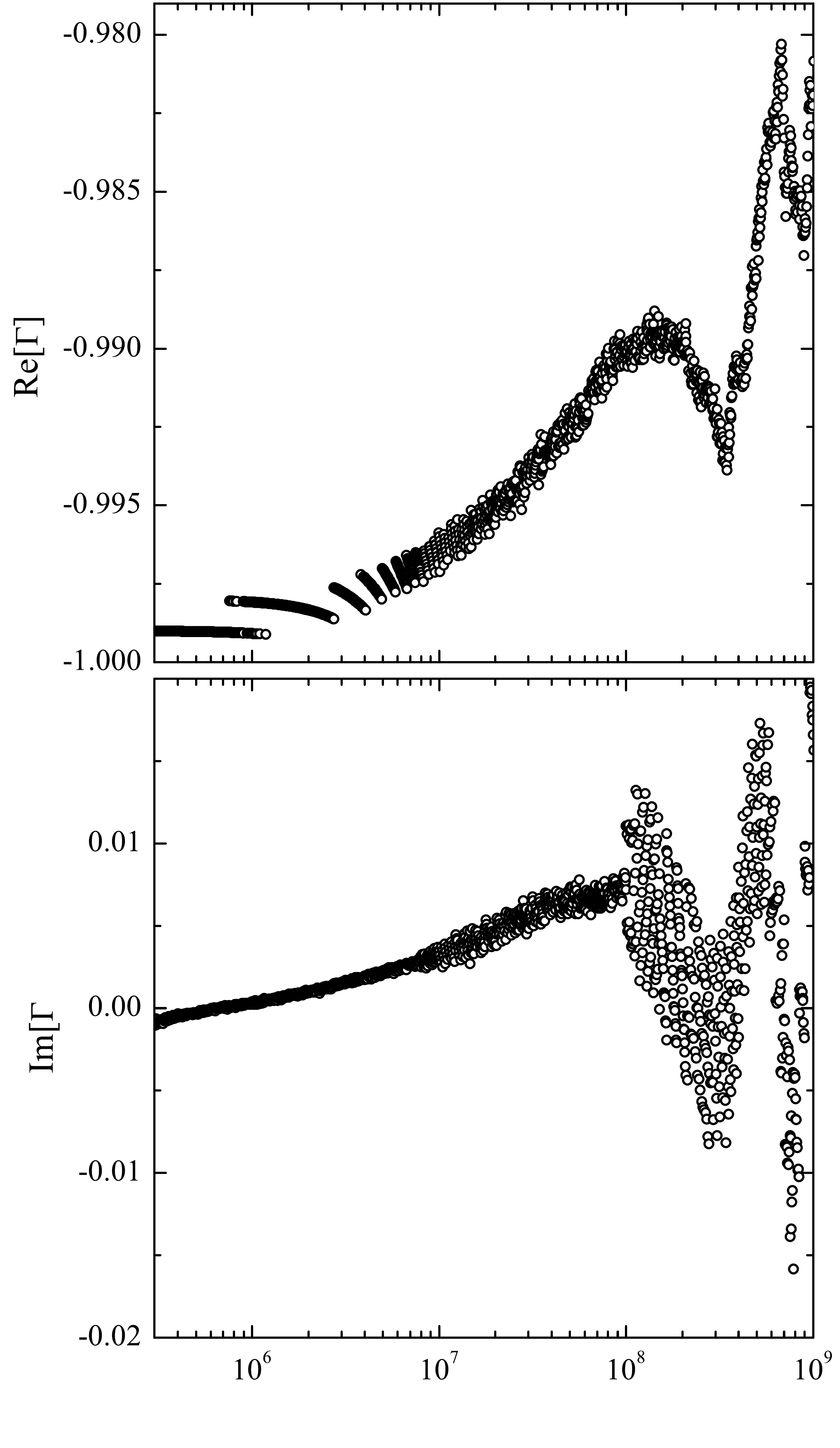}
\caption{\label{fig:short}The real (top) and imaginary (bottom) components of $\Gamma(\omega)$ of the short-circuited UT-141 coaxial probe as determined from the measured $\rho_{m}(\omega)$ using Eq.~\ref{eq:Gamma2} and the fitted value of $\beta\ell$.  Note the very fine scales of the vertical axes.  The step-like feature of $\mathrm{Re}[\Gamma]$ at low frequency is due to a digitization of the VNA measurement of $\mathrm{Re}[\rho_{m}]$.}
\end{center}
\end{figure}

As a specific example, the low-frequency measurements made in this work were obtained using an Agilent E5061A 300~kHz to 1.5~GHz vector network analyzer and a probe constructed from a 20.7~cm length of \mbox{UT-141} semirigid coaxial cable with silver-plated copperweld center conductor, copper outer conductor, and polytetrafluoroethylene dielectric.  A calibration plane was established at the connector of the probe using the Agilent 85033E calibration kit.  All measurements taken using the Agilent E5061A VNA were made using an incident power of 10~dBm and a log-frequency sweep with 1601 frequency points.  With the open end of the probe short circuited, fits to Eq.~\ref{eq:rho} yielded \mbox{$\sqrt{\varepsilon}\ell/c=0.981\pm0.001$~ns} which is close to the expected value assuming $\varepsilon=2.2$.  We report the asymptotic standard error of the nonlinear least-squares fit as the uncertainty of the fit parameter $\sqrt{\varepsilon}\ell/c$.  Unless specified otherwise, all reported uncertainties are standard errors associated with a least-squares fit or derived from standard errors using a propagation of errors analysis.  As a final confirmation that the data analysis procedure was reliable, Fig.~\ref{fig:short} shows the real and imaginary parts of $\Gamma(\omega)$ of the short-circuited probe from 300~kHz to 1~GHz as determined from $\rho_{m}(\omega)$ using Eq.~\ref{eq:Gamma2}.
Up to 100~MHz, the short-circuited probe behaved as a nearly ideal transmission line such that $0.99<\left\vert\Gamma\right\vert<1$ and \mbox{$\left\vert\mathrm{Im}[\Gamma]/\mathrm{Re}[\Gamma]\right\vert<0.007$}.

\subsection{Determining $C_{f}$ and $C_0$}\label{sec:CfC0}

At frequencies below the onset of any dispersion of the MUT permittivity, $\varepsilon^\prime$ is nearly constant while $\varepsilon^{\prime\!\prime}$ is negligible.  In this limit the load impedance, as given by Eq.~\ref{eq:Zinv}, can be reexpressed as:
\begin{equation}
Z_{L}=\frac{R^{-1}-j\omega C_{T}}{R^{-2}+(\omega C_{T})^2}\label{eq:impedance}
\end{equation}
where $R^{-1}=C_0\sigma_{dc}/\varepsilon_0$ and $C_{T}=C_{f}+\varepsilon^\prime C_0$.  If, for a particular MUT, $\sigma_{dc}=0$ the impedance reduces to $Z_{L}=1/j\omega C_{T}$.

\begin{figure}[tb]
\begin{center}
\includegraphics[width=.75\columnwidth]{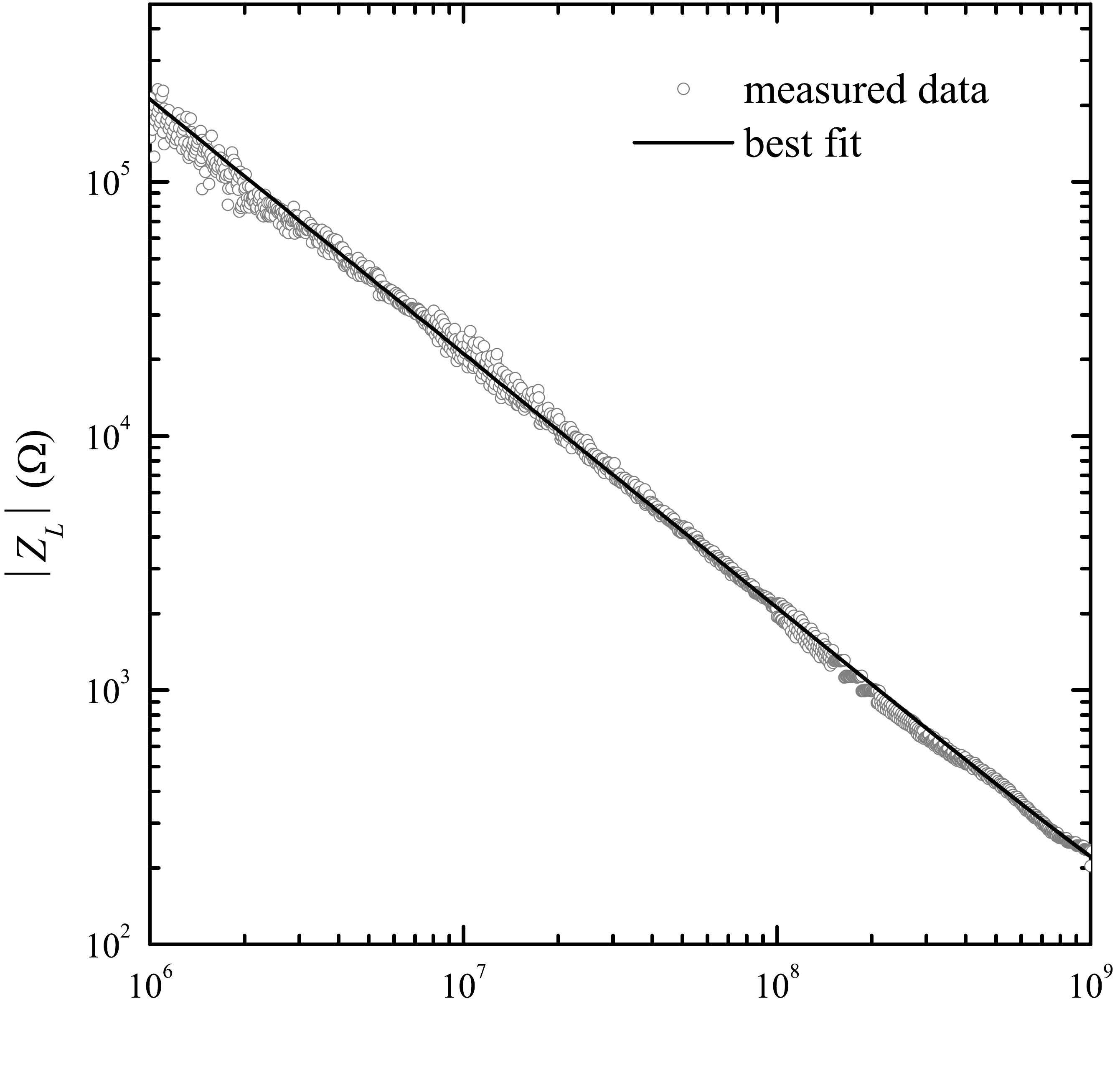}
\caption{\label{fig:meth}The magnitude of $Z_{L}$ for a 99.9\% methyl alcohol sample.  The data have been fitted to $1/\omega C_{T}$ from 3.0~MHz to 1.0~GHz yielding $C_{T}=0.752\pm0.001$~pF.}
\end{center}
\end{figure}

The impedance of a particular probe can be characterized using two samples of known $\varepsilon_{r}$.  For example, to characterize the \mbox{UT-141} probe, 99.9\% methyl alcohol at $28.0\pm0.1^\circ$C and a solution of 99.8\% NaCl dissolved in water at $25.0\pm0.1^\circ$C were used.  The sample temperatures were set using a commercial temperature-controlled water bath.  The resolution to the temperature regulation was 0.1$^\circ$C.  Methyl alcohol has negligible conductivity and Bao and coworkers give a Deybe-type dispersion for its permittivity~\cite{Bao:1996}:
\begin{equation}
\varepsilon_{r}^\mathrm{meth}=\varepsilon_{h}+\frac{\Delta\varepsilon}{1+j\omega\tau}\label{eq:methPerm}
\end{equation}
where the parameters at 28$^\circ$C are \mbox{$\varepsilon_{h}=6.6\pm0.4$}, \mbox{$\Delta\varepsilon=26.7\pm0.7$}, and \mbox{$\tau=52.6\pm 1.7$~ps}.  Provided \mbox{$\omega\ll\tau^{-1}$}, to a very good approximation \mbox{$\varepsilon_{r}^\mathrm{meth}$} is purely real and equal to \mbox{$\varepsilon_{h}+\Delta\varepsilon=33.3\pm0.8$}.  Figure~\ref{fig:meth} shows $\left\vert Z_{L}\right\vert$ of the methyl alcohol sample as determined from a measurement of $\rho_m$ and Eqs.~\ref{eq:Gamma1} and \ref{eq:Gamma2}.
These data were fitted to \mbox{$\left\vert Z_{L}\right\vert=1/\omega C_{T}$} and the best fit curve is shown in the figure.

\begin{figure}[htb]
\begin{center}
\includegraphics[width=.77\columnwidth]{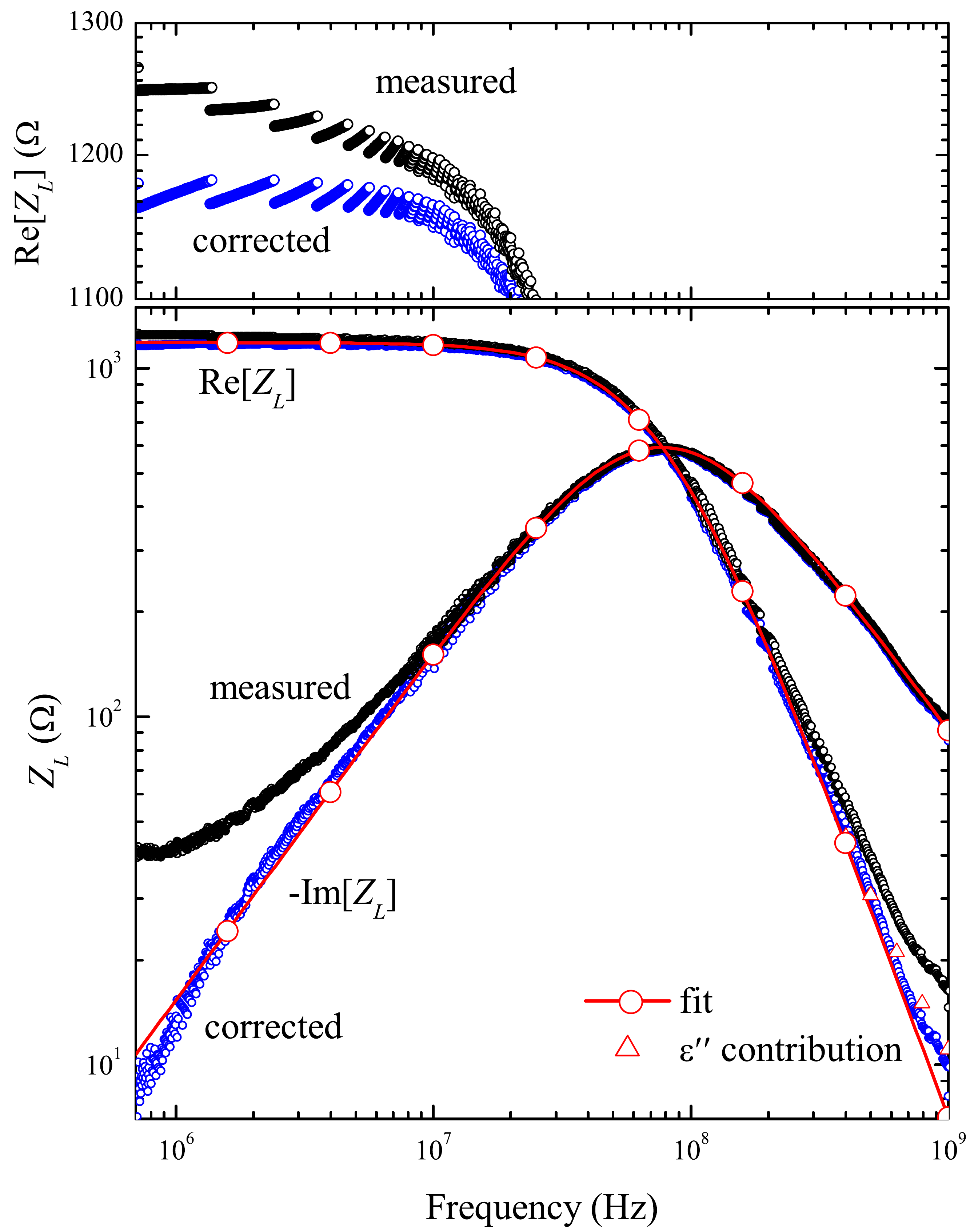}
\caption{\label{fig:NaCl}The black data points show the real and imaginary parts of the measured $Z_{L}$ for a 0.03 molarity solution of NaCl in water.  The data have been fitted to Eq.~\ref{eq:impedance} from 20 to 300~MHz.  The deviations from the expected behavior at low frequency are due to the electrode polarization effect discussed in Sec.~\ref{sec:electrodePolarization}.  Above 300~MHz, the deviations in $\mathrm{Re}[Z_{L}]$ are due to weak electrode polarization and the onset of dispersion in $\varepsilon_{r}$ of water.  The open triangles show the additional contribution to $\mathrm{Re}\left[Z_{L}\right]$ expected from the dispersion.  The blue data sets show the real and imaginary components of the corrected load impedance after removing the electrode polarization contributions.  The top frame highlights differences between the measured and corrected $\mathrm{Re}\left[Z_{L}\right]$ data sets at low frequencies.}
\end{center}
\end{figure}

A 0.03 molarity solution of 99.8\% NaCl dissolved in distilled water was used as the second control sample.  The distilled water was further purified using the PURELAB$\textregistered$ Ultra water purification system to achieve a minimum resistivity of 18.2~M$\Omega$-cm.  In Ref.~\cite{Stogryn:1971}, Stogryn gives formulae for calculating the dc conductivity of NaCl in water as a function of molarity and temperature.  These formulae give $\sigma_{dc}=0.31~\Omega^{-1}\mathrm{m}^{-1}$ for a 0.03 molarity solution at 25$^\circ$C.  Buchner {\it et al.\/} have determined the complex permittivity of pure water between 0 and 35$^\circ$C~\cite{Buchner:1999},  at low frequencies and 25$^\circ$C, $\varepsilon_{r}^{\mathrm{H}_2\mathrm{O}}$ is approximately constant and equal to 78.32. Figure~\ref{fig:NaCl} shows the effective impedance of the probe tip when it is submerged in the NaCl solution. The data span a frequency range of 700~kHz to 1~GHz and were determined from a measurement of $\rho_m$ and Eqs.~\ref{eq:Gamma1} and \ref{eq:Gamma2}.
The data have been fitted to Eq.~\ref{eq:impedance} resulting in best-fit parameters $R=1187\pm 14~\Omega$ and $C_{T}=1.73\pm 0.06$~pF.

Below 10~MHz, the measured data deviate from the model expectations due to the electrode polarization effect.  The mechanisms responsible for electrode polarization and a reliable method for removing its effects are presented in Sec.~\ref{sec:electrodePolarization}.  Above 300~MHz, the real part of $Z_{L}$ deviates from the modeled behavior in part due to diminished, but lingering, electrode polarization effects and in part to the breakdown of the assumption that $\varepsilon_{r}^{\mathrm{H}_2\mathrm{O}}$ is real and constant.  The open triangles in Fig.~\ref{fig:NaCl} show the additional contribution to $\mathrm{Re}\left[Z_{L}\right]$ expected from the onset of $\varepsilon_{r}^{\mathrm{H}_2\mathrm{O}}$ dispersion.

Having found $C_{T}=C_{f}+\varepsilon^\prime C_0$ for two distinct samples, $C_{f}$ and $C_0$ can in principle be independently determined.  For the UT-141 transmission line, this analysis resulted in \mbox{$C_0=0.0217\pm0.0014$~pF} and \mbox{$C_{f}=0.029\pm0.089$~pF}.  The large uncertainty in $C_{f}$ is expected since $C_{T}\approx\varepsilon_{r}C_0$ when $\varepsilon_{r}\gg 1$.  In this case, approximate uncertainties of $\delta C_0\sim\delta C_{T}/\varepsilon_{r}$ and $\delta C_{f}\sim \sqrt{2}\,\delta C_{T}$ are anticipated.  Using Eqs.~\ref{eq:Zinv} and \ref{eq:Gamma1} it is straightforward to show that the electrical properties of the MUT are related to the reflection coefficient $\Gamma$ via:
\begin{equation}
\varepsilon^\prime-j\left(\varepsilon^{\prime\!\prime}+\frac{\sigma_{dc}}{\omega\varepsilon_0}\right)=\frac{1}{j\omega Z_0 C_0}\left(\frac{1-\Gamma}{1+\Gamma}\right)-\frac{C_{f}}{C_0}.\label{eq:extract}
\end{equation}
This expression makes it immediately clear that $C_{f}$ only affects the real part of the permittivity.  Moreover, for biological samples, one is typically in the limit \mbox{$\varepsilon^\prime\gg 1$} over the entire measurement bandwidth of the open-ended coaxial probe and, since $C_{f}/C_0$ is of order one or less, the measured permittivity is insensitive to the value of $C_{f}$.

Finally, the fitted value of $R$ and the value of $C_0$ can be used to calculate experimental values for the dc conductivity of the NaCl solution.  The result was \mbox{$\sigma_{dc}=0.34\pm 0.02~\Omega^{-1}\mathrm{m}^{-1}$} which is close to the expected value of \mbox{$0.31~\Omega^{-1}\mathrm{m}^{-1}$}~\cite{Stogryn:1971}.

To summarize, at low frequencies the coaxial probe can be approximated as a length ideal transmission line terminated by an effective impedance that is dependent on the electrical properties of the material surrounding its tip.  The measured reflection coefficient $\rho_m$ at the calibration plane of the VNA is related to the reflection coefficient $\Gamma$ at the probe tip via Eq.~\ref{eq:Gamma2}.  Once the shunt capacitances $C_0$ and $C_f$ of a particular probe are known, the permittivity and conductivity of the MUT are determined using Eq.~\ref{eq:extract}.  Both $\varepsilon^{\prime\!\prime}$ and $\sigma_{dc}$ are independent of $C_f$ and $\varepsilon^\prime$ is usually very insensitive to its value.

\section{High-Frequency Analysis}\label{sec:analysis}

At sufficiently high frequencies the coaxial probe can no longer be reliably approximated as an ideal transmission line and corrections are required to account for spurious reflections and for ohmic and dielectric losses.  For example, Fig.~\ref{fig:short} shows that the short-circuited \mbox{UT-141} probe exhibits nonideal behavior at frequencies above 100~MHz.  The high-frequency analysis requires more sophisticated methods to calculate $\Gamma$ from $\rho_m$ measurements.  In this section, we treat the probe as a two-port network and use a scattering matrix to relate the two reflection coefficients. The elements $S_{ij}$ of the scattering matrix are found by terminating the open end of the probe with known loads.  Determining the $S_{ij}$, however, requires that the shunt capacitances at the open end of the probe are known.  In subsection~\ref{sec:Amethod}, an alternative calibration method that does not require specific knowledge of these capacitances is described.  Finally, the permittivity of the methyl alcohol control sample is measured to verify that the analysis techniques are reliable and to expose their limitations.

\subsection{Scattering Matrix Calibration}
\begin{figure}[tb]
\begin{center}
\includegraphics[width=.65\columnwidth]{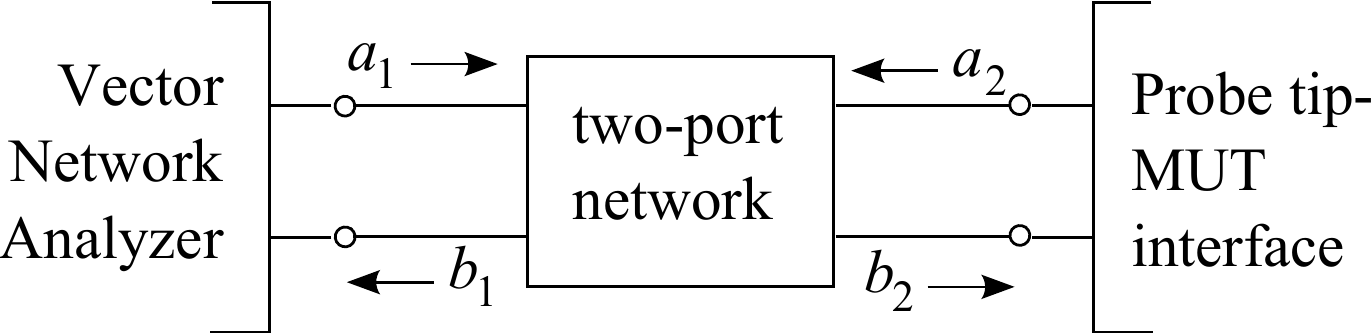}
\caption{\label{fig:2port}A general two-port network with signals incident ($a_i$) and reflected ($b_i$) from ports $i=1$ and $2$.}
\end{center}
\end{figure}

The calibration plane of the VNA can be extended to the open end of the coaxial probe by treating the probe as a two-port network~\cite{Bao:1994}.  A general two-port network is as shown in Fig.~\ref{fig:2port}.
Port 2 is the probe tip-MUT interface and port 1 is connected to the VNA test port cable.  The incident and reflected signals $a_i$ and $b_i$ are related via the scattering matrix~\cite{Collin:2001}:
\begin{equation}
\left[ {\begin{array}{c}
b_1 \\
b_2 \\
\end{array} }\right]
 =
\left[ {\begin{array}{cc}
 S_{11} & S_{12}  \\
 S_{21} & S_{22}  \\
 \end{array} } \right]
 \left[ {\begin{array}{c}
a_1 \\
a_2 \\
\end{array} }\right]\label{eq:matrix}
\end{equation}
where $S_{ij}$ are the scattering parameters.  The ratio of the signals reflected from and incident on port 1 is the measured reflection coefficient \mbox{$\rho_{m}=b_1/a_1$} while the desired reflection coefficient at port 2 is given by \mbox{$\Gamma=a_2/b_2$}.  When combined with Eq.~\ref{eq:matrix}, these definitions of the reflection coefficients can be used to solve for $\Gamma$ in terms of $\rho_{m}$:
\begin{equation}
\Gamma=\frac{\rho_{m}-S_{11}}{S_{22}\rho_{m}+S_{12}S_{21}-S_{11}S_{22}}.\label{eq:cal}
\end{equation}
The unknown scattering parameters are determined by terminating port 2 with known loads $Z_{{L},i}$ for which $\Gamma_i$ can be calculated and $\rho_i$ can be measured.  To independently determine each of $S_{11}$, $S_{22}$, and the product $S_{12}S_{21}$, three standard loads ($i=1$, 2, and 3) must be measured.  In terms of $\Gamma_i$ and $\rho_i$, the scattering parameters are:
\begin{align}
S_{11} &=\frac{\rho_3 T_{12}+\rho_2 T_{31}+\rho_1 T_{23}}{T_{12}+T_{31}+T_{23}}\\
S_{22}&=\frac{\Gamma_1(\rho_2-S_{11})+\Gamma_2(S_{11}-\rho_{1})}{\Gamma_1\Gamma_2(\rho_{2}-\rho_{1})}\\
S_{12}S_{21}&=\frac{(\rho_{1}-S_{11})(1-S_{22}\Gamma_1)}{\Gamma_1}\label{eq:S12S21}
\end{align}
where the quantity $T_{ij}$ has been defined as:
\begin{equation}
T_{ij}\equiv \Gamma_i\Gamma_j\left(\rho_i-\rho_j\right).
\end{equation}
In a typical calibration, short-circuit, open circuit, and matched load terminations are used~\cite{Kraszewski:1983, daSilva:1978, Wei:1989}.  For the coaxial probe, a reliable short-circuit termination is achieved by pressing the probe tip into a thin indium sheet and an open-circuit termination is achieved by suspending the probe tip in free space.  However, the probe cannot be easily terminated with a broadband matched-load.  In its place, a standard liquid load of known permittivity is used as the third termination.  The known reflection coefficients are easily calculated using:
\begin{equation}
\Gamma_i=\frac{Z_{{L},i}-Z_0}{Z_{{L},i}+Z_0}\label{eq:standGamma}
\end{equation}
where the impedances for the short- and open-circuit terminations are given by \mbox{$Z_{{L},1}=0$} and \mbox{$Z_{{L},2}=1/j\omega(C_{f}+C_0)$} respectively, while the standard liquid impedance $Z_{{L},3}$ is given by Eq.~\ref{eq:Zinv}.

Once the scattering parameters of a particular probe are known, a measurement of $\rho_m$ and Eqs.~\ref{eq:extract} and \ref{eq:cal} are used to determine the unknown permittivity and conductivity of a MUT.

\subsection{Alternative Calibration Method}\label{sec:Amethod}

Bao and coworkers developed a calibration scheme that uses the same load terminations, but does not require specific knowledge of the values of $C_{f}$ and $C_0$~\cite{Bao:1994}.  The strategy is to use Eqs.~\ref{eq:Zinv} and \ref{eq:Gamma1} to express $\Gamma$ in terms of $\varepsilon_{r}-j\sigma_{dc}/\omega\varepsilon_0$ and then to equate it to Eq.~\ref{eq:cal}.  A rearrangement of the resulting expression yields:
\begin{equation}
\rho_{m}=\frac{A_2+A_3\left(\varepsilon_{r}-j\frac{\sigma_{dc}}{\omega\varepsilon_0}\right)}{A_1+\left(\varepsilon_{r}-j\frac{\sigma_{dc}}{\omega\varepsilon_0}\right)}\label{eq:A}
\end{equation}
where:
\begin{align}
A_1&=\frac{1-S_{22}}{j\omega Z_0 C_0(1+S_{22})}+\frac{C_{f}}{C_0}\\
A_2&=\frac{S_{11}-S_{11}S_{22}+S_{12}S_{21}}{j\omega Z_0 C_0(1+S_{22})}\nonumber\\
&~~~~~+\frac{S_{11}+S_{11}S_{22}-S_{12}S_{21}}{1+S_{22}}\frac{C_{f}}{C_0}\\
A_3&=\frac{S_{11}+S_{11}S_{22}-S_{12}S{21}}{1+S_{22}}.
\end{align}
The advantage of this analysis is that $A_1$, $A_2$, and $A_3$ can be determined purely from the measured reflection coefficients of the standard terminations.  For the short-circuit termination, $\Gamma_1=-1$, and Eq.~\ref{eq:cal} immediately leads to $A_3=\rho_1$.  For the open-circuit ($\varepsilon_{r}=1$, $\sigma_{dc}=0$) and standard liquid terminations, Eq.~\ref{eq:A} gives:
\begin{align}
\rho_2&=\frac{A_2+A_3}{A_1+1}\\
\rho_3&=\frac{A_2+A_3\left(\varepsilon_{r}^{s}-j\frac{\sigma_{dc}^{s}}{\omega\varepsilon_0}\right)}{A_1+\left(\varepsilon_{r}^{s}-j\frac{\sigma_{dc}^{s}}{\omega\varepsilon_0}\right)}
\end{align}
where $\varepsilon_{r}^{s}$ and $\sigma_{dc}^{s}$ are the relative permittivity and dc conductivity of the standard liquid load. These expressions can be solved for the two unknowns $A_1$ and $A_2$ such that:
\begin{align}
A_1&=\frac{(\rho_2-\rho_1)+(\rho_1-\rho_3)\left(\varepsilon_{r}^{s}-j\frac{\sigma_{dc}^{s}}{\omega\varepsilon_0}\right)}{\rho_3-\rho_2}\label{eq:A1}\\
A_2&=\frac{\rho_3(\rho_2-\rho_1)+\rho_2(\rho_1-\rho_3)\left(\varepsilon_{r}^{s}-j\frac{\sigma_{dc}^{s}}{\omega\varepsilon_0}\right)}{\rho_3-\rho_2}\label{eq:A2}\\
A_3&=\rho_1.\label{eq:A3}
\end{align}
With $A_1$, $A_2$, and $A_3$ all in terms of known or measurable quantities, a measurement of $\rho_{m}$ for any MUT and  Eq.~\ref{eq:A} are all that are needed to completely determine an unknown $\varepsilon_{r}-j\sigma_{dc}/\omega\varepsilon_0$.

\begin{figure}[htb]
\begin{center}
\includegraphics[width=.75\columnwidth]{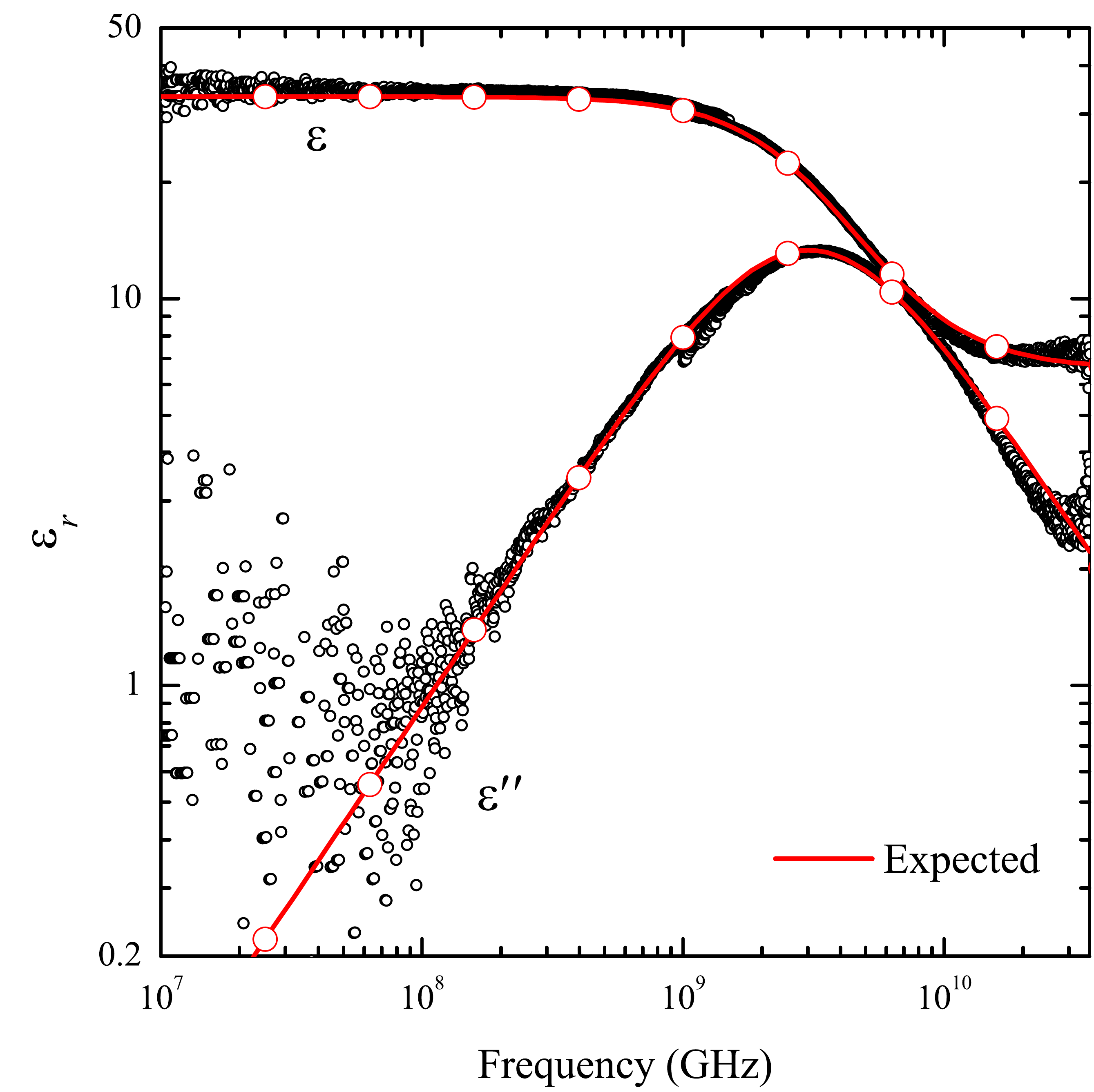}
\caption{\label{fig:methPerm}The real and imaginary parts of $\varepsilon_{r}$ of a 99.9\% sample of methyl alcohol at $28^\circ$C from 10~MHz to 36~GHz.  The solid curves show the expected permittivity and were generated using Eq.~\ref{eq:methPerm} with parameters obtained from Ref.~\cite{Bao:1996}.}
\end{center}
\end{figure}

To demonstrate the capabilities of the open-ended coaxial probe, Fig.~\ref{fig:methPerm} shows permittivity data of a 99.9\% pure methyl alcohol sample.
Below 1.5~GHz, data were acquired using an Agilent E5061A 300~kHz to 1.5~GHz rf network analyzer with the Flexco Microwave FC195 cable test set and an open-ended coaxial probe made from \mbox{UT-141} semirigid transmission line.  Above 1.5~GHz, data were acquired using an Agilent 8722ES 50~MHz to 40~GHz microwave vector network analyzer with the Agilent 85133F cable test set and an open-ended probe made from \mbox{UT-085} semirigid transmission line.  In both cases, a calibration plane was established at the connector of the probe using the Agilent 85033E calibration kit and data were collected using log-frequency sweeps with 1601 frequency points.  We confirmed that all of our results were independent of the incident signal power.  All of the data presented here were taken using the maximum signal power (10~dBm for the Agilent E5061A and -10~dBm for the Agilent 8722ES).  The data shown in the figure were analyzed using Eq.~\ref{eq:A} and Eqns.~\ref{eq:A1} -- \ref{eq:A3}.  Pure water was used as the third calibration standard.  For $\varepsilon_{r}$ of pure water, a superposition of two Debye processes, as given by Buchner {\it et al.\/} in Ref.~\cite{Buchner:1999} was used:
\begin{equation}
\varepsilon_{r}^{\mathrm{H}_2\mathrm{O}}=\frac{\varepsilon-\varepsilon_2}{1+j\omega\tau_1}+\frac{\varepsilon_2-\varepsilon_\infty}{1+j\omega\tau_2}+\varepsilon_\infty\label{eq:H2Operm}
\end{equation}
where at 25$^\circ$C the parameters are $\varepsilon=78.32$, $\tau_1=8.38$~ps, $\varepsilon_2=6.32$, $\tau_2=1.1$~ps, and $\varepsilon_\infty=4.57$.
When the data are analyzed using the $S$-parameter method (Eq.~\ref{eq:extract} and Eqns.~\ref{eq:cal} - \ref{eq:S12S21}) the results for $\varepsilon_{r}$ of the MUT are identical.

The measured permittivity data in Fig.~\ref{fig:methPerm} deviate from the expected $\varepsilon_{r}^\mathrm{meth}$ above 30~GHz.  This systematic error results from evanescent modes excited at the probe-MUT junction when the cutoff frequencies of higher order modes are exceeded.  The simplest way to extend the measurement frequency range is to construct a probe from smaller diameter coaxial cable (UT-047, for example).  Alternatively, methods that can be used to correct for evanescent modes and radiation effects are given in Refs.~\cite{Bao:1996}, \cite{Athey:1982}, and \cite{Stuchly:1994}.  For biological materials, it is the permittivity below 30~GHz that is of primary interest and we make no attempt to model these high-frequency effects.

For a sample with negligible dc conductivity (such as methyl alcohol) the lower end of the measurement frequency range is set, not by extrinsic effects, but by limitations on the measurement resolution.  This fact is most easily seen by separating Eq.~\ref{eq:extract} into its real and imaginary components:
\begin{align}
\varepsilon^\prime &= \frac{1}{\omega Z_0 C_0}\left(\frac{-2\Gamma^{\prime\!\prime}}{1+2\Gamma^\prime+\vert\Gamma\vert^2}\right)-\frac{C_{f}}{C_0}\\
\varepsilon^{\prime\!\prime} &=\frac{1}{\omega Z_0 C_0}\left(\frac{1-\vert\Gamma\vert^2}{1+2\Gamma^\prime+\vert\Gamma\vert^2}\right)
\end{align}
where $\sigma_{dc}$ has been set to zero and \mbox{$\Gamma\equiv\Gamma^\prime+j\Gamma^{\prime\!\prime}$}.  This analysis shows that \mbox{$\varepsilon^\prime\propto\Gamma^{\prime\!\prime}/\left(\omega Z_0 C_0\right)$} and \mbox{$\varepsilon^{\prime\!\prime}\propto\left(1-\vert\Gamma\vert^2\right)/\left(\omega Z_0 C_0\right)$}.  At sufficiently low frequencies \mbox{$\omega Z_0 C_0\ll 1$} and \mbox{$\Gamma\approx\Gamma^\prime\approxeq 1$} such that both $\varepsilon^\prime$ and $\varepsilon^{\prime\!\prime}$ are evaluated from the ratio to two quantities that are both approaching zero.  Probes made from larger diameter coaxial cable will have higher shunt capacitance and could be used to extend the bottom end of the measurement bandwidth.

\section{Electrode Polarization}\label{sec:electrodePolarization}

Measurements of the low-frequency permittivity of samples with appreciable dc conductivity will be contaminated with contributions from the electrode polarization effect.  Care must be taken to either use experimental \mbox{setups} that mitigate these effects or to use reliable data correction schemes to remove them post-measurement~\cite{Schwan:1968, Schwan:1992, Kuang:1998, Bordi:2001}.

\begin{figure}[t]
\begin{center}
\includegraphics[width=.6\columnwidth]{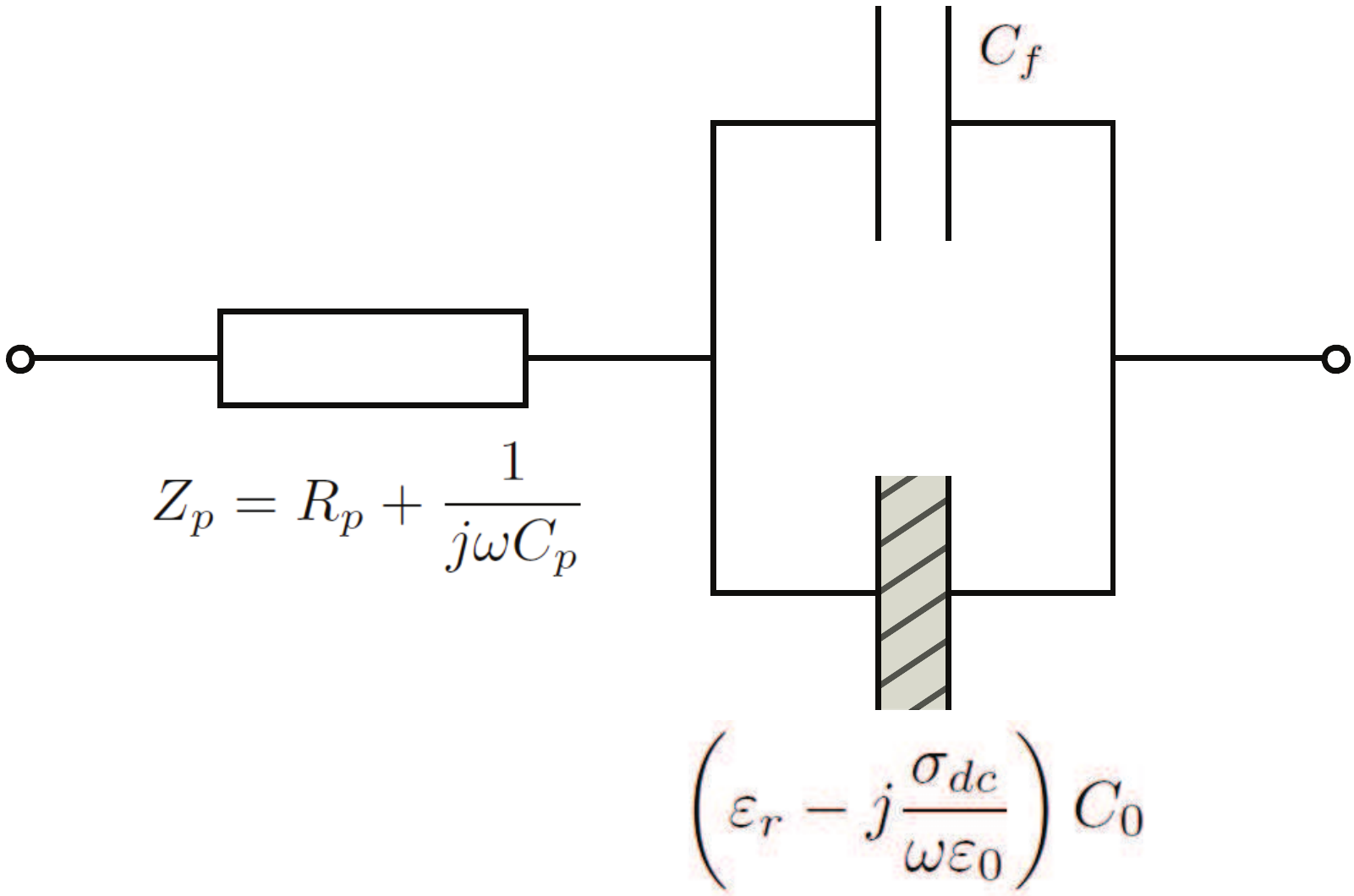}
\caption{\label{fig:electrodeZ}Electrode polarization effects can be modeled with a complex impedance $Z_{p}$ in series with the intrinsic impedance due to the submerged tip of the open-ended coaxial probe.  The MUT, of complex relative permittivity $\varepsilon_{r}$ and conductivity $\sigma_{dc}$, is represented by the hatched region filling capacitor $C_0$.}
\end{center}
\end{figure}

When submerged in an electrolytic solution, a metallic electrode will acquire a surface charge due, for example, to dissociation of electrode surface molecules or to the absorption of ions from the solution.  In response to this surface charge, the concentration of oppositely charged ions from the electrolyte increases in the region of the electrode.  These counterions are effectively bound and form a so-called electrical double layer at the electrode-electrolyte interface.  This separation of charge can be modeled as a capacitor in series with the effective impedance of the probe-MUT interface.  In practice, due to electrochemical reactions at the electrode, it is typically necessary to include a conductance term such that an accurate model necessitates a full complex electrode polarization impedance $Z_{p}=R_{p}+1/j\omega C_{p}$.  For the open-ended coaxial probe, Fig.~\ref{fig:electrodeZ} shows the effective load impedance present at the probe tip when electrode polarization effects are included.
Low-frequency permittivity measurements are challenging as the electrode polarization impedance is dependent on the measurement frequency, sample conductivity, electrode geometry, and current density~\cite{Kuang:1998}.

One way to significantly reduce the effects of electrode polarization, without modifying the experimental geometry, is to coat the electrodes with platinum.  The platinum coating does not readily react with the electrolytic solution thereby reducing the surface charge acquired by the electrode and hence the electrode polarization impedance.  An additional coating of platinum-black (pt-black) can further suppress the polarization impedance by orders of magnitude as the porous coating greatly enhances the effective surface area of the electrodes~\cite{Schwan:1968}.  When determining the permittivity of a MUT, it is necessary to clean the tip of the probe while alternating between measurements of calibration samples and the MUT.  Mechanical or ultrasonic scrubbing of the probe tip would cause the pt-black coating to degrade making the calibration measurements unreliable.

Rather than modify the coaxial probe or the experimental geometry, the methods of Raicu {\it et al.\/} can be followed to reliably remove the electrode polarization artifacts from the data post-measurement~\cite{Raicu:1998}.  Through careful experimentation it has been found that the real and imaginary components of the electrode polarization impedance can be modeled using power laws \mbox{$R_{p}=A\left(\omega/1~\mathrm{rad~s}^{-1}\right)^{-m}$} and \mbox{$C_{p}=B\left(\omega/1~\mathrm{rad~s}^{-1}\right)^{-n}$} where $m$ and $n$ are positive constants less than one~\cite{Schwan:1992, Raicu:1998}.  The powers $m$ and $n$ are related through the Kramers-Kronig relations for the polarization impedance and are expected to obey Fricke's law: \mbox{$m+n=1$}~\cite{Schwan:1992,Fricke:1932,Asami:1993}.  When combined with Eq.~\ref{eq:impedance}, the net load impedance at low frequencies can be written as:
\begin{align}
Z_{L}=&A\left(\frac{\omega}{1~\mathrm{rad/s}}\right)^{-m}+\frac{R}{1+\left(\omega R C_{T}\right)^2}\nonumber\\
&-j\left[\frac{(1~\mathrm{rad/s})^{m-1}}{B\omega^{m}}+\frac{\omega R^2 C_{T}}{1+\left(\omega R C_{T}\right)^2}\right].\label{eq:30}
\end{align}
The parameters $A$, $B$, and $m$ can be determined by making use of the fact that:
\begin{align}
-\frac{d\mathrm{Re}\left[Z_{L}\right]}{d\omega}=&\frac{mA\omega^{-m-1}}{(1~\mathrm{rad/s})^{-m}}+\frac{2\omega R^3 C_{T}^2}{\left[1+\left(\omega R C_{T}\right)^2\right]^2}\label{eq:dReA}\\
\frac{d\mathrm{Im}\left[Z_{L}/\omega\right]}{d\omega}=&\frac{(m+1)(1~\mathrm{rad/s})^{m-1}}{B\omega^{m+2}}+\frac{2\omega R^4 C_{T}^3}{\left[1+\left(\omega R C_{T}\right)^2\right]^2}.\label{eq:dImA}
\end{align}
For sufficiently low frequencies such that $\omega R C_{T}\ll 1$, the frequency dependencies of both of the above derivatives are dominated by the electrode polarization terms.

\begin{figure}[tb]
\begin{center}
\includegraphics[width=.7\columnwidth]{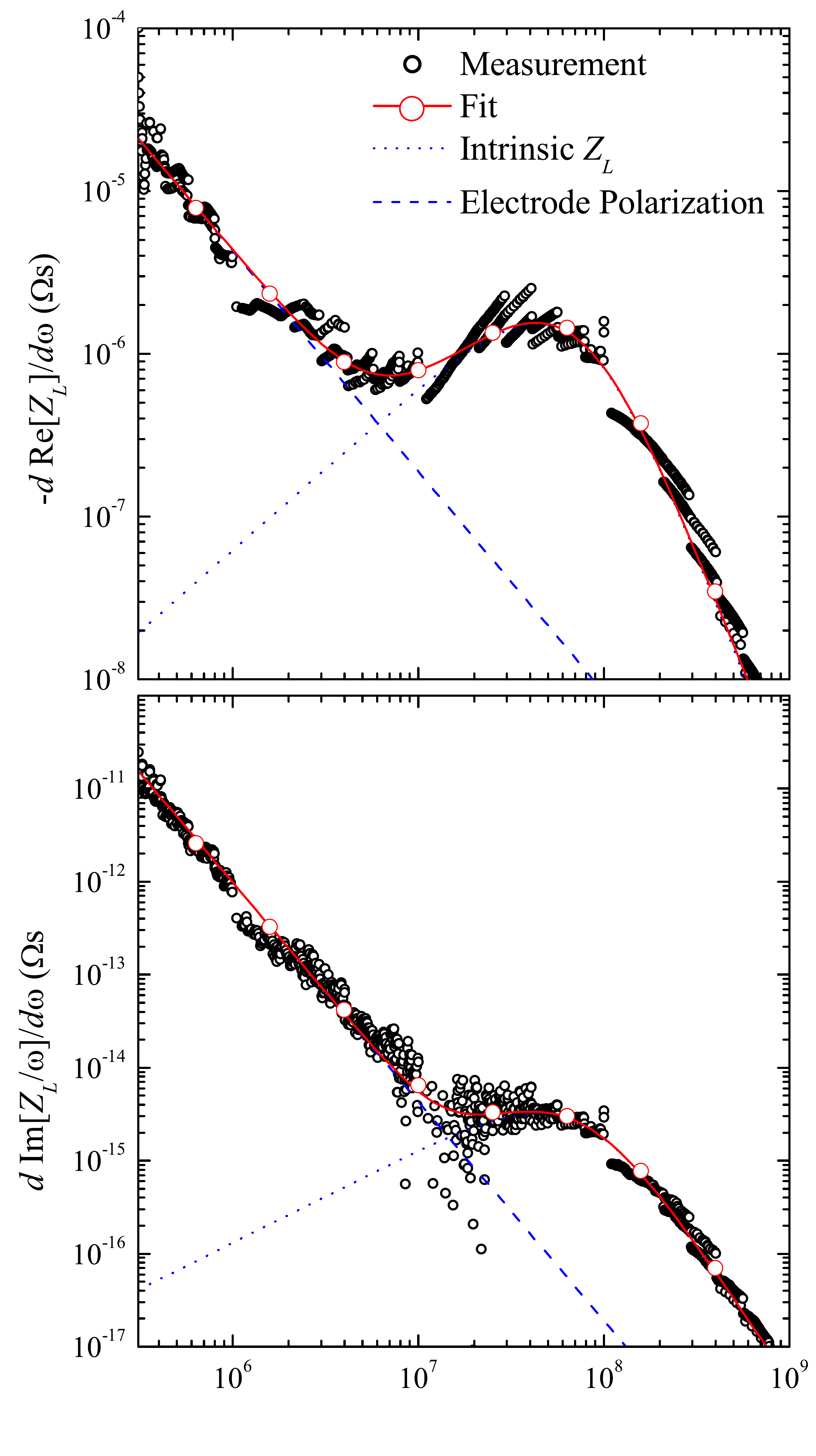}
\caption{\label{fig:correction}Logarithmic plots of $-d\mathrm{Re}\left[Z_{L}\right]/d\omega$ (top) and $d\mathrm{Im}\left[Z_{L}/\omega\right]/d\omega$ (bottom) as a function of frequency for the aqueous NaCl solution.  The low-frequency power law behaviors can be used to parameterize $R_{p}$ and $C_{p}$.  The solid curves are fits to Eqs.~\ref{eq:dReA} and \ref{eq:dImA}.  The dashed lines highlight the electrode polarization contributions.}
\end{center}
\end{figure}

To demonstrate the viability of this correction scheme, the method was applied to the measured impedance data of the 0.03~molarity NaCl solution previously shown in Fig.~\ref{fig:NaCl}.  For this sample, the relevant time constant is \mbox{$\tau=R C_{T}=2.05\pm0.08$~ns} such that electrode polarization effects are expected to dominate for frequencies much less than \mbox{$1/2\pi\tau=75$~MHz}.  Figure~\ref{fig:correction} shows $-d\mathrm{Re}\left[Z_{L}\right]/d\omega$ and $d\mathrm{Im}\left[Z_{L}/\omega\right]/d\omega$ for the NaCl sample as a function of frequency as determined from the measured impedance.
The anticipated low-frequency power law behavior due to electrode polarization is clearly exhibited below 1~MHz as shown by the dashed lines.  Simultaneous fits to Eqs.~\ref{eq:dReA} and \ref{eq:dImA} yielded the parameters \mbox{$m=0.356\pm0.019$}, \mbox{$A=20\pm 5~\mathrm{k}\Omega$}, and \mbox{$B=130\pm 40~\upmu\mathrm{F}$}.
Electrode polarization effects were removed by subtracting $R_{p}+1/j\omega C_{p}$ from the measured load impedance.  Figure~\ref{fig:NaCl} shows $Z_{L}$ of the NaCl sample both before and after the correction was applied.  For both the real and imaginary components of $Z_{L}$, the correction scheme removed the effects of electrode polarization and the corrected data exhibited the expected low-frequency behavior shown by the solid curves.  Note also that the correction scheme adjusted the high-frequency tail of $\mathrm{Re}\left[Z_{L}\right]$ such that it properly shows the subtle effect of the $\varepsilon_{r}$ dispersion.

\begin{figure}[t]
\begin{center}
\includegraphics[width=.75\columnwidth]{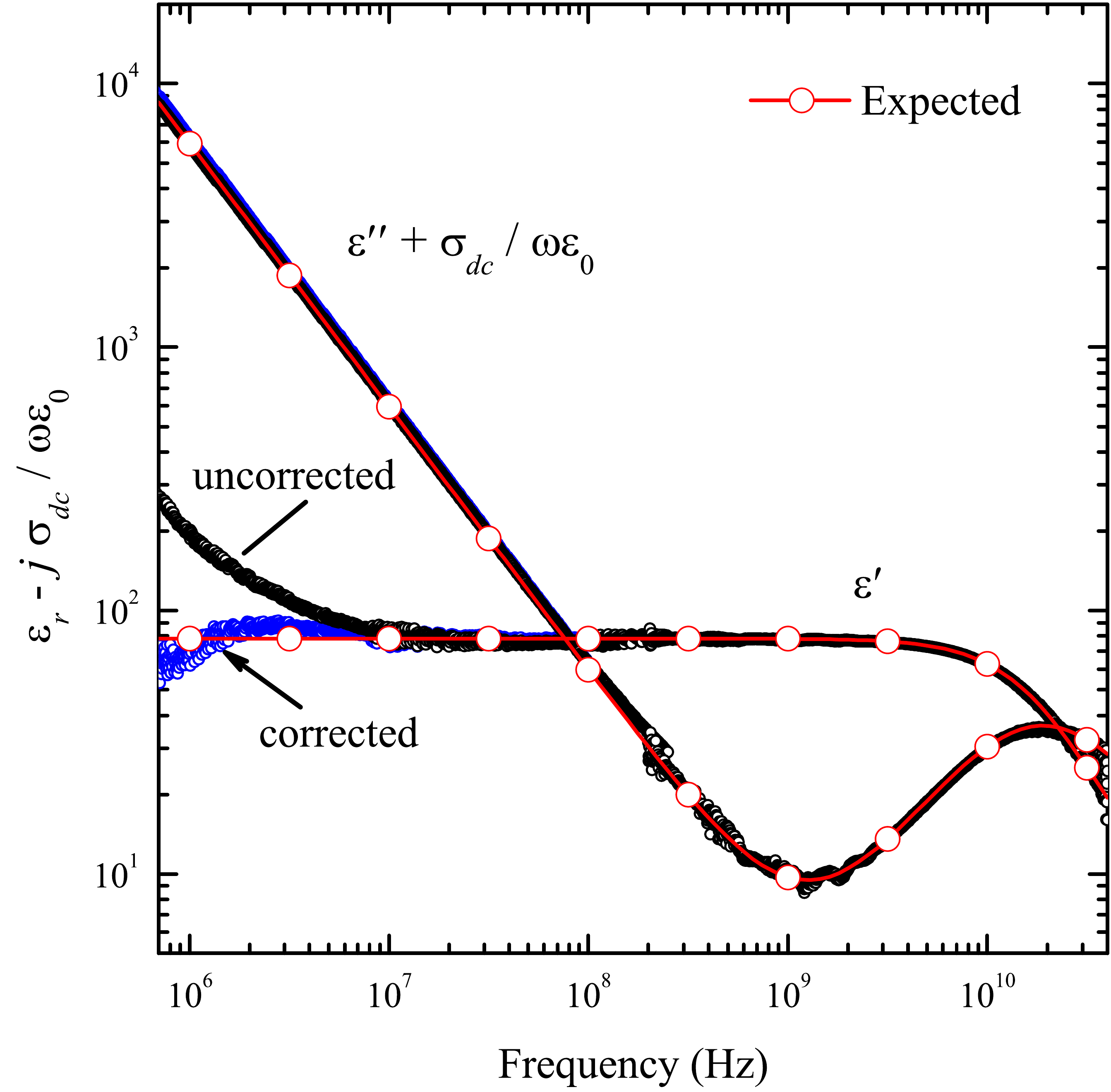}
\caption{\label{fig:NaClPerm}The real and imaginary parts of \mbox{$\varepsilon_{r}-j\sigma_{dc}/\omega\varepsilon_0$} of a 0.03 molarity solution of NaCl dissolved in pure water.  The sample temperature was regulated at $25^\circ$C.  The solid curves show the expected behavior and were generated using $\varepsilon_{r}$ of water and $\sigma_{dc}=0.33~\Omega^{-1}\mathrm{m}^{-1}$.  At low frequency, the results are shown both before (black) and after (blue) the electrode polarization correction was applied.  The correction has no effect on the imaginary component of the data, but the upturn in the real data set is an artifact of electrode polarization.}
\end{center}
\end{figure}

Figure~\ref{fig:NaClPerm} shows the measured permittivity and conductivity of the NaCl solution from 700~kHz to 40~GHz at 25$^\circ$C.
To show the effect of electrode polarization, $\varepsilon_{r}-j\sigma_{dc}/\omega\varepsilon_0$ has been determined both before and after applying the correction scheme.  At low frequencies, the imaginary component of the data is dominated by the conductivity term and the corrected and uncorrected $\varepsilon^{\prime\!\prime}+\sigma_{dc}/\omega\varepsilon_0$ data sets are indistinguishable. On the other hand, the low-frequency upturn of $\varepsilon^\prime$ is an artifact of electrode polarization that is suppressed in the corrected data. The solid lines in Figs.~\ref{fig:NaClPerm} are not free-parameter fits to the data, but are predictions made using the permittivity of pure water and a conductivity of $\sigma_{dc}=0.33~\Omega^{-1}\mathrm{m}^{-1}$ which is consistent with the value previously determined in Sec.~\ref{sec:CfC0}.

The electrode polarization correction scheme described in this section is straightforward to implement, but has limited range of applicability.  As shown in Figs.~\ref{fig:NaCl} and \ref{fig:NaClPerm}, the correction scheme extended the lower end of our measurement bandwidth by an order of magnitude.  As the frequency is reduced further, the polarization impedance $Z_p$  becomes much larger than the intrinsic impedance of the sample and the correction scheme ceases to be reliable.  We conclude this section with a very brief discussion of two alternative experimental geometries that can be used to significantly suppress electrode polarization effects~\cite{ Kaatze:2006, Schwan:1992}.

The first method makes use of a pair of electrodes, usually in a parallel plate geometry, whose separation distance can be manipulated. The time constant $RC_T$ in Eq.~\ref{eq:30} is independent of the distance between electrodes while $R$ is proportional to the separation distance.  Increasing the distance between electrodes therefore suppresses electrode polarization contribution to $Z_L$ and allows for reliable measurements at lower frequencies.
The second geometry makes use two pairs of electrodes.  One pair supplies current to the MUT while the other pair senses the potential difference across the sample.  Provided that the sense electrodes draw negligible current, by using a high-impedance amplifier for example, the electrode polarization effect will be suppressed and low-frequency frequency measurements can be made reliably~\cite{ Kaatze:2006, Schwan:1992}.

\section{WAS Permittivity}\label{sec:WAS}

\begin{figure}[tb]
\begin{center}
\includegraphics[width=.68\columnwidth]{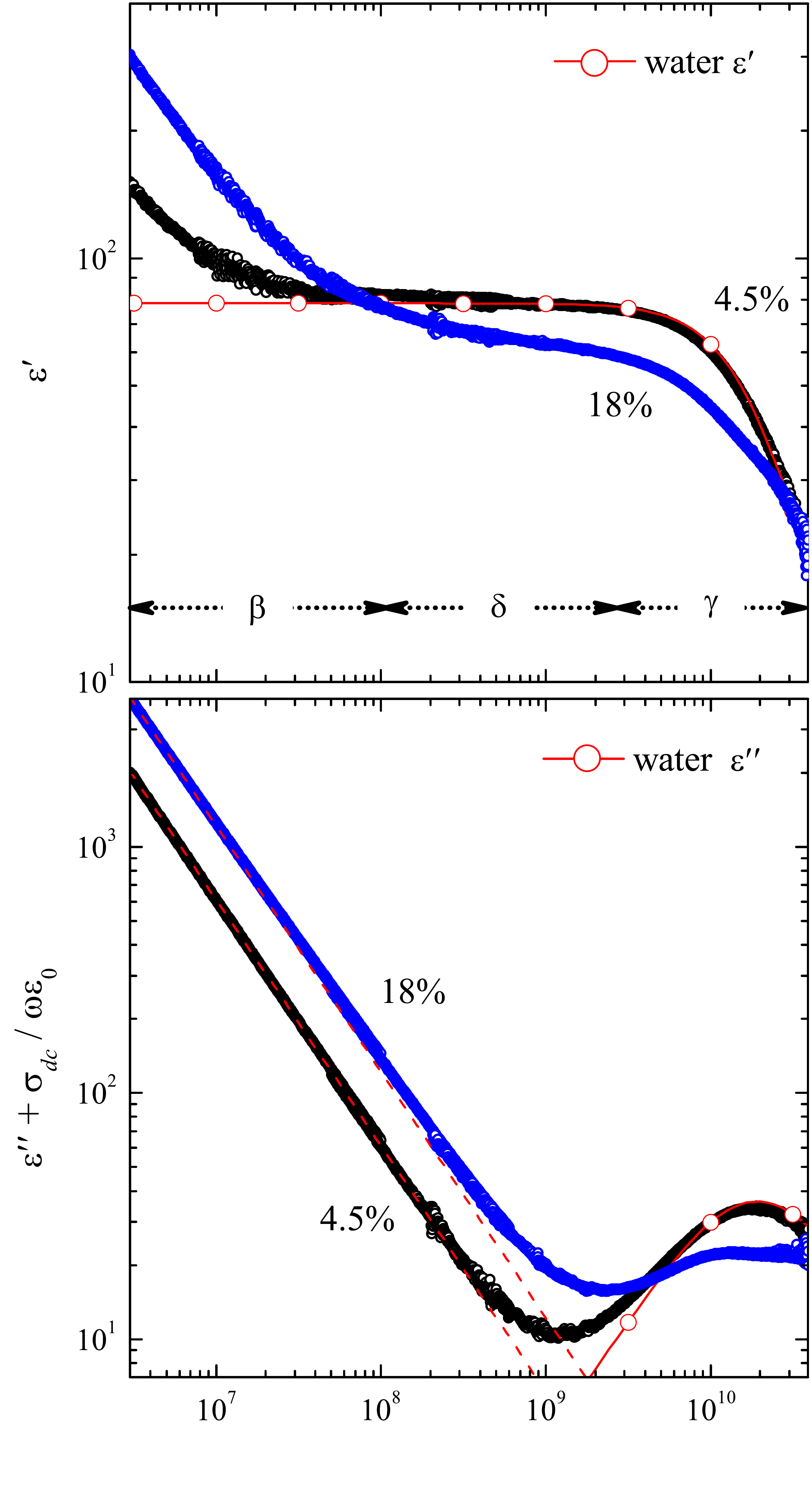}
\caption{\label{fig:WASperm}The real (top) and imaginary (bottom) components of $\varepsilon_{r}-j\sigma_{dc}/\omega\varepsilon_0$ of WAS as a function of frequency from 3~MHz to 40~GHz.  Samples with both 4.5\% (black) and 18\% (blue) solid concentrations were measured.  The dashed lines have slope -1 and represent $\sigma_{dc}/\omega\varepsilon_0$ contributions.  The solid red curve is $\varepsilon_r$ of pure water.}
\end{center}
\end{figure}

Having demonstrated the feasibility of the open-ended coaxial probe using methyl alcohol and NaCl solution control samples, the system was next used to determine the unknown permittivity of WAS obtained from our local WWTF in Kelowna, British Columbia, Canada.  WAS is comprised of different groups of microorganisms, organic and inorganic matter agglomerated together in a polymeric network formed by microbial extracellular polymeric substances (EPS) and cations~\cite{Li:1990}.  Here, we present the experimental data to highlight the capabilities of the measurement technique, but refer readers to a previous publication for a detailed discussion of the results~\cite{Bobowski:2012}.  Two concentrations of WAS are common at WWTFs.  One, referred to as thickened WAS (TWAS), contains 4.5\% solids by weight.  Before final disposal, TWAS is routinely dewatered using a centrifuge resulting in a material called ``sludge cake'' that is 18\% solid by weight.  Figure~\ref{fig:WASperm} shows the real and imaginary components of $\varepsilon_{r}-j\sigma_{dc}/\omega\varepsilon_0$ for both the 4.5\% and 18\% WAS samples after correcting for electrode polarization effects.

Short-circuit, open-circuit, and the 0.03 molarity NaCl solution were used as calibration standards.  The NaCl solution was chosen because its permittivity is well known and mimics some essential features of the expected electrical properties of the WAS samples.  We emphasize that, below $10$~MHz, electrode polarization effects become non-negligible and the NaCl solution ceases to be a reliable calibration standard. It is, therefore, necessary to use the methods of Sec.~\ref{sec:probe} to analyze the low-frequency data.  One might consider using a standard liquid having negligible $\sigma_{dc}$, like pure water or methyl alcohol, to avoid electrode polarization during calibration. However, for these liquids $\varepsilon^{\prime\!\prime}\ll \varepsilon^\prime$ at low frequency and the effective impedance at the probe tip, as given be Eq.~\ref{eq:Zinv}, becomes $Z_{L,3}\approx 1/j\omega\left(C_f+\varepsilon^\prime C_0\right)$.  As a result, $\Gamma_2$ of the open-circuit probe and $\Gamma_3$ of the standard liquid termination are both real and approach one as frequency is lowered (see Eq.~\ref{eq:standGamma}).  Without three independent standard terminations, the calibration techniques presented in Sec.~\ref{sec:analysis} are not effective and one must make use of the data analysis methods presented in Sec.~\ref{sec:probe}.

For the results shown in Fig.~\ref{fig:WASperm}, data below 100~MHz were acquired using a \mbox{UT-141} probe and analyzed using the methods of Sec.~\ref{sec:probe} and data above 50~MHz were acquired using a \mbox{UT-085} probe and analyzed using the methods of Sec.~\ref{sec:analysis}. The 50~MHz region of overlap was used to confirm that consistent results were obtained using both the low-frequency and high-frequency analysis techniques.  Finally, the scheme described in Sec.~\ref{sec:electrodePolarization} was used to remove the electrode polarization effects from the low-frequency data.

As is typical of a wide variety of biological samples~\cite{Pethig:1984}, the permittivity of WAS can be subdivided into distinct regions.  Above 100~MHz, $\varepsilon^\prime$ of 4.5\% WAS mimics the permittivity of pure water.  This dispersion, referred to as $\gamma$-dispersion, is due to the relaxation of polar water molecules.  At low frequencies, $\varepsilon^\prime$ of both the 4.5\% and 18\% samples show upturns.  This effect, called $\beta$-dispersion, is caused by the Maxwell-Wagner effect in which charge accumulates at cell walls which separate the intra- and extracellular fluids of the sample~\cite{Pethig:1984, Kuang:1998}.  The 18\% sample exhibits an additional weak dispersion at intermediate frequencies from 100~MHz to 10~GHz.  This effect, generally called $\delta$-dispersion, is attributed to the relaxation of water molecules bound to adjacent proteins~\cite{Foster:1981, Foster:1980}.

The lower plot of Fig.~\ref{fig:WASperm} shows that $\varepsilon^{\prime\!\prime}+\sigma_{dc}/\omega\varepsilon_0$ of both samples is dominated by the conductivity term below 1~GHz.  The dashed lines in the figure follow a $\omega^{-1}$ frequency dependence and are drawn using $\sigma_{dc}=0.34~\Omega^{-1}\mathrm{m}^{-1}$ and $0.68~\Omega^{-1}\mathrm{m}^{-1}$ for the 4.5\% and 18\% samples respectively.  Above a few gigahertz, the conductivity term is negligible and the data are dominated by $\varepsilon^{\prime\!\prime}$.  As expected, the 4.5\% sample tracks the $\gamma$-dispersion of bulk water very closely, whereas the 18\% sample shows significant deviations due to an active $\delta$-dispersion.

\section{Summary}\label{sec:conclusions}

Open-ended coaxial transmission lines and vector network analyzers were used to make reliable measurements of the complex permittivity and dc conductivity of liquid and semisolid samples over a broad frequency range.  The load impedance $Z_L$ at the probe tip was modeled as parallel shunt capacitances $C_f$ and $C_0$. Capacitor $C_0$ is immersed in the MUT and allows $\varepsilon_r$ and $\sigma_{dc}$ of the material to be experimentally determined.

The low-frequency data were analyzed by treating the coaxial probe as an ideal transmission line.  This analysis technique was verified using a methyl alcohol control sample.  For materials having appreciable dc conductivity, the electrode polarization effect modifies the effective impedance at the probe tip.  By including a polarization impedance $Z_p$ in series with $Z_L$, the effects of electrode polarization were reliably characterized and removed post-measurement as was demonstrated using a second control sample of NaCl dissolved in water.  The correction scheme extended the bottom end of our measurement bandwidth by more than an order of magnitude.

Above a few hundred megahertz, the coaxial probe can no longer be considered an ideal transmission line.  Rather, it is necessary to model the probe as a two-port network.  A calibration using open-circuit, short-circuit, and standard liquid terminations was implemented to correct for spurious reflections and losses that occur beyond the calibration plane of the VNA.  The calibration procedure was experimentally verified using the methyl alcohol and the aqueous NaCl control samples.

Having fully characterized the open-ended coaxial probe, it was next used to determine the complex permittivity and dc conductivity of waste-activated sludge.  Samples having both 4.5\% and 18\% solid concentrations were measured.  In both samples, $\varepsilon^\prime$ showed an upturn below 100~MHz due to the charging of cell membranes. This $\beta$-dispersion was enhanced in the 18\% due to a higher concentration of suspended cells.  At high frequency, $\varepsilon^\prime$ of the 4.5\% sample mimics that of pure water.  The 18\% sample shows an additional weak dispersion at intermediate frequencies which is due to the increased relaxation times of polar water molecules bound to adjacent proteins.

\ack
The financial support of the Natural Science and Engineering Research Council of
Canada Strategic Project Grant (\#396519-10) is gratefully acknowledged.  The Agilent 8722ES VNA used in this work was provided by CMC Microsystems and their support is gratefully acknowledged.


\begin{thebibliography}{99}

%1
\bibitem{Stuchly:1982} Stuchly,~M.~A., T.~Whit~Athey, G.~M.~Samaras, and G.~E.~Taylor, ``Measurement of radio frequency permittivity of biological tissues with an open-ended coaxial line: part II -- experimental results," {\it IEEE Trans. Microwave Theor. Techn.\/}, Vol.~MIT--30, No.~1, 87--92, 1982.

%2
\bibitem{Foster:1981} Foster,~K.~R. and J.~L.~Schepps, ``Dielectric properties of tumor and normal tissues at radio through microwave frequencies," {\it J. Microwave Power\/}, Vol.~16, No.~2, 107--119, 1981.

%3
\bibitem{Bao:1996} Bao,~J.-Z., M.~L.~Swicord, and C.~C.~Davis, ``Microwave dielectric characterization of binary mixtures of water, methanol, and ehtanol," {\it J. Chem. Phys.\/}, Vol.~104, No.~12, 4441--4450, 1996.

%4
\bibitem{Erle:2000} Erle,~U., M.~Regier, C.~Persch, and H.~Schubert, ``Dielectric properties of emulsions and suspensions: mixture equations and measurement comparisons," {\it J. Microw. Power Electromagn. Energy\/}, Vol.~35, No.~3, 185--190, 2000.

%5
\bibitem{Wang:2003} Wang,~Y., T.~D.~Wig, J.~Tang, and L.~M.~Hallberg, ``Dielectric properties of foods relevant to rf and microwave pasteurization and sterilization," {\it J. Food Eng.\/}, Vol.~57, No.~3, 257--268, 2003.

%6
\bibitem{El-Rayes:1987} El-Rayes,~M.~A. and F.~T.~Ulaby, ``Microwave dielectric spectrum of vegetation--part 1: experimental observations," {\it IEEE Trans. Geosci. Remote Sensing\/}, Vol.~GE--25, No.~5, 541--549, 1987.

%7
\bibitem{Jackson:1990} Jackson,~T.~J., ``Laboratory evaluation of a field-portable dielectric/soil-moisture probe," {\it IEEE Trans. Geosci. Remote Sensing\/}, Vol.~28, No.~2, 241--245, 1990.
%8
\bibitem{Kaatze:2006} Kaatze,~U. and Y.~Feldman, ``Broadband dielectric spectrometry of liquids and biosystems," {\it Meas. Sci. Technol.\/}, Vol.~17, No.~2, R17--R35, 2006.

%9
\bibitem{Li:1990} Li,~D.~H. and J.~J.~Ganczarczyk, ``Structure of activated sludge floes," {\it Biotechnol. Bioeng.\/}, Vol.~35, No.~1, 57--65, 1990.

%10
\bibitem{Bobowski:2012} Bobowski,~J.~S., T.~Johnson, and C.~Eskicioglu, ``Permittivity of waste-activated sludge by an open-ended coaxial line," {\it Prog. Electromagn. Res. Lett.\/} Vol.~29, 129--139, 2012.

%11
\bibitem{Eskicioglu:2009} Eskicioglu,~C., K.~J.~Kennedy, and R.~L.~Droste, ``Enhanced disinfection and methane production from sewage sludge by microwave irradiation," {\it Desalination\/}, Vol.~248, No.~1--3, 279--285, 2009.

%12
\bibitem{Appels:2008} Appels,~L., J.~Baeyens, J.~Degr\`eve, and R.~Dewil, ``Principles and potential of the anaerobic digestion of waste-activated sludge," {\it Prog. Energy Combust. Sci.\/}, Vol.~34, No.~6, 755--781, 2008.

%13
\bibitem{Pethig:1984} Pethig,~R., ``Dielectric properties of biological materials: biophysical and medical applications," {\it IEEE Trans. Electr. Insul.\/}, Vol.~{EI--19}, No.~5, 453--474, 1984.

%14
\bibitem{Stuchly:1980} Stuchly,~M.~A. and S.~S.~Stuchly, ``Coaxial line reflection methods for measuring dielectric properties of biological substances at radio and microwave frequencies -- a review," {\it IEEE Trans. Instrum. Meas.\/}, Vol.~IM--29, No.~3, 176--183, 1980.

%15
\bibitem{Jarvis:1994} Baker-Jarvis,~J., M.~D.~Janezic, P.~D.~Domich, and R.~G.~Geyer, ``Analaysis of an open-ended coaxial probe with lift-off for nondestructive testing," {\it IEEE Trans. Instrum. Meas.\/}, Vol.~43, No.~5, 711--718, 1994.

%16
\bibitem{Stogryn:1971} Stogryn,~A., ``Equations for calculating the dielectric constant of saline water," {\it IEEE Trans. Microwave Theor. Techn.\/}, Vol.~19, No.~8, 733--736, 1971.

%17
\bibitem{Buchner:1999} Buchner,~R., J.~Barthel, and J.~Stauber, ``The dielectric relaxation of water between $0^\circ$C and $35^\circ$C," {\it Chem. Phys. Lett.\/}, Vol.~306, No.~1--2, 57--63, 1999.

%18
\bibitem{Bao:1994} Bao,~J.-Z., C.~C.~Davis, and M.~L.~Swicord, ``Microwave dielectric measurements of erythrocyte suspensions," {\it Biophys. J.\/}, Vol.~66, No.~6, 2173--2180, 1994.

%19
\bibitem{Collin:2001} Collin,~R.~E., {\it Foundations for Microwave Engineering, $2^{nd}$ Ed.}, John Wiley \& Sons, New Jersey, 2001.

%20
\bibitem{Kraszewski:1983} Kraszewski,~A., M.~A.~Stuchly, and S.~S.~Stuchly, ``ANA calibration method for measurements of dielectric properties," {\it IEEE Trans. Instrum. Meas.\/}, Vol.~IM--32, No.~2, 385--387, 1983.

%21
\bibitem{daSilva:1978} da~Silva,~E.~F. and M.~K.~McPhun, ``Calibration techniques for one port measurement," {\it Microwave J.\/}, Vol.~21, No.~6, 97--100, 1978.

%22
\bibitem{Wei:1989} Wei,~Y.-Z. and S.~Sridhar, ``Technique for measuring the frequency-dependent complex dielectric constants of liquids up to 20 GHz," {Rev. Sci. Instrum.\/}, Vol.~60, No.~9, 3041--3046, 1989.

%23
\bibitem{Athey:1982} Whit~Athey,~T., M.~A.~Stuchly, and S.~S.~Stuchly, ``Measurement of radio frequency permittivity of biological tissues with an open-ended coaxial line: part I," {\it IEEE Trans. Microwave Theor. Techn.\/}, Vol.~MIT--30, No.~1, 82--86, 1982.

%24
\bibitem{Stuchly:1994} Stuchly,~S.~S., C.~L.~Sibbald, and J.~M.~Anderson, ``A new aperture admittance model for open-ended waveguides," {\it IEEE Trans. Microwave Theor. Techn.\/}, Vol.~42, No.~2, 192--198, 1994.

%25
\bibitem{Schwan:1968} Schwan,~H.~P., ``Electrode polarization impedance and measurements in biological materials," {\it Ann. New York Acad. Sci.\/}, Vol.~148, No.~1, 191--209, 1968.

%26
\bibitem{Schwan:1992} Schwan,~H.~P., ``Linear and nonlinear electrode polarization and biological materials," {\it Ann. Biomed. Eng.\/}, Vol.~20, No.~3, 269--288, 1992.

%27
\bibitem{Kuang:1998} Kuang,~W. and S.~O.~Nelson, ``Low-frequency dielectric properties of biological tissues: a review with some new insights," {\it Trans. ASAE\/}, Vol. 41, No.~1, 173--184, 1998.

%28
\bibitem{Bordi:2001} Bordi,~F, C.~Cametti, and T.~Gili, ``Reduction of the contribution of electrode polarization effects in the radiowave dielectric measurements of highly conductive biological cell suspensions," {Bioelectrochemistry\/}, Vol.~54, No.~1, 53--61, 2001.

%29
\bibitem{Raicu:1998} Raicu,~V., T.~Saibara, and A.~Irimajiri, ``Dielectric properties of rat liver in vivo: a noninvasive approach using an open-ended coaxial probe at audio/radio frequencies," {\it Bioelectrochem. Bioenerg.\/}, Vol.~47, No.~2, 325--332, 1998.

%30
\bibitem{Fricke:1932} Fricke, H., ``The theory of electrolytic polarization," {\it Phil. Mag.\/}, Vol.~14, No.~90, 310--318, 1932.

%31
\bibitem{Asami:1993} Asami,~K. and T.~Hanai, ``Observations and the phenomenological interpretation of dielectric relaxation due to electrode polarization," {\it Bull. Inst. Chem. Res. Kyoto Univ.\/}, Vol.~71, No.~2, 111--119, 1993.

%32
\bibitem{Foster:1980} Foster,~K.~R., J.~L.~Schepps, and H.~P.~Schwan, ``Microwave dielectric relaxation in muscle. A second look," {\it Biophys. J.\/}, Vol.~29, No.~2, 271--281, 1980.




\end{thebibliography}
\end{document}